\documentclass[aps,prd,twoside,twocolumn,floatfix,letterpaper]{revtex4}

\usepackage[utf8]{inputenc}
\usepackage[dvipdfmx]{graphicx}

\usepackage{latexsym}

\usepackage{pdfpages}
\usepackage{subfig}
\usepackage{float}
\usepackage{textcomp}
\usepackage{amssymb}
\usepackage{array}
\usepackage{amsmath,amssymb}
\usepackage[T1]{fontenc}
\usepackage{graphics}
\usepackage{bm}
\usepackage{pifont}
\newcommand{\hcond}{\usefont{T1}{phv}{mc}{n}}
\makeatletter
\def\section{\@startsection{section}{1}{0pt}{-3.5ex plus -1ex minus -.2ex}{2.3ex plus .2ex}{\large \hcond}}
\def\subsection{\@startsection{subsection}{2}{\z@}{-3.25ex plus -1ex minus -.2ex}{1.5ex plus .2ex}{\hcond}}

\begin{document}

\setcounter{page}{1}

%\newcommand{\angstrom}{\textup{\AA} }

%\title{Problem of an electrically charged quantum particle on the axis of symmetry of an electrostatically charged ring}

\title{Quantum problem of the potential of a ring charged on the symmetry axis}

\author{Wytler Cordeiro dos Santos}\email[]{wytler.cordeiro@unb.br}
\author{Bruno Carmo Nunes} \email[]{brunocmo.nunes@gmail.com}
\author{Ronni G.G. Amorim}\email[]{ronniamorim@gmail.com}
\affiliation{Faculty of Gama, University of Bras\'ilia - UnB, 72444-240, Gama-DF, Brazil.}

\date{\today}

\begin{abstract}
%Propomos um problema de Mecânica Quântica de uma partícula com carga elétrica colocada sobre o eixo de simetria de um anel carregado eletricamente em coordenadas cilíndricas. Calculamos numericamente alguns dos autovalores de %
In this work we discuss about the problem of an electrically charged particle placed on the symmetry axis of an electrically charged ring in a quantum viewpoint. This problem should be an expanded version of the usual quantum ring and quantum corral. For this purpose, we present a detailed and pedagogical review about a version of the quantum ring focusing in mathematical aspects. As a new result, we calculate numerically the spectrum and wave functions related to charged particle located on the symmetry axis of charged ring and respective wave functions using two different ways: Numerov and perturbative method.
\end{abstract}

\maketitle

\section{Introduction}
%\indent

The Solid State Physics became a very relevant subject from the end of the 1930's  with the formulation of the theory of the structure of electronic bands. The limiting condition for the validity of this theory is the assumption of an infinitely extended crystal without defects, interfaces or surfaces \cite{Kittel}. Approximately in the same time, the Quantum Mechanics problem of a particle in a one-dimensional potential well was solved for the first time.
From 1947, with the development of the transistor by Shockley and collaborators, many other devices presented the decisive milestones for the advancement of technologies that allowed the creation and development of semiconductors for solar cells, microprocessors and semiconductor based lasers \cite{Kittel}.
In the development of Solid State Physics, nanostructures should be highlighted. Important examples of nanostructures are carbon nanotubes, quantum wires, conductive polymers, nanocrystalline semiconductors, metallic nanoparticles and quantum dots \cite{Kittel}. A simple example of a quantum dot is an electron in a spherical potential well, that is, an electron confined in a rigid spherical shell. It is important to emphasize that inside the spherical shell the electron is free. Under this boundary condition this problem has been solved analytically \cite{Griffths}.
%%%%%%%%%%%%%%%%%%%%%%%%%%%%%%%%%%%%%%%%%%
%%%%%%%%%%%%%%%%%%%%%%%%%%%%%%%%%%%%%%%%%%%

In the last two decades, nanoestructures of type quantum rings have gained prominence  due to the potential interest for nano-devices \cite{chakraborty, barticevic}. The main physical idea is a ring nanostructure (one-dimensional circle) with a few units or hundreds of charge carriers, such as electrons distributed in the nanostructure of the ring. The number of charge carriers is well defined and its fluctuation  has to be under experimental control \cite{Danilevich}.
In a quantum ring one must observe and measure the interaction of the ring with an external magnetic field. If the magnetic field is restricted to a circular region with a radius smaller than the radius of the ring, the physical interaction that arises is the Aharonov-Bohm effect \cite{Fomin}.
In this reference, the author presents as the initial idea of quantum rings a Pauli hypothesis that to calculate the magnetically induced current densities in the aromatic hydrocarbon ring molecules it must be considered that the external electrons in the benzene molecule can circulate freely and provide a very large contribution to the diamagnetic susceptibility with the magnetic field normal to the hexagonal plane of carbon atoms.

Quantum rings present novel magnetic and optoelectronic properties; a subject matter that has attracted an increasing interest in several areas including materials physics and engineering. Potential applications of quantum rings are THz detectors \cite{browmick, huang}, solar cells \cite{wu}, and lasers \cite{suarez}. Recently, quantum rings theory was applied in study of novel quantum materials with superconducting qubits \cite{neill} as well as in the study of the electronic properties of graphene \cite{jellal}. In the present work we analyze a structure similar to quantum rings. We discuss the nontrivial problem of a charged particle located on the symmetry axis of a quantum ring. In this sense, the energy spectrum and wave functions are determined by using a perturbative approach and the so-called Numerov method.  It is relevant to mention that the system analyzed in this work had not yet been studied in the literature. Thus, despite being a theoretical approach, it may bring important experimental applications.

The presentation of this paper is organized in the following structure. In Section II,  in order to fix the notation and to review the problem,  a theoretical discussion of a quantum corral addressed previously \cite{Eigler1,Eigler2, Eigler3} is presented. We use the Schrödinger equation to analytically calculate the allowable quantum states for a quantum particle confined to the quantum corral. This theoretical model of the quantum corral is fundamental to discuss the problem proposed in this article. In Section III, we present the mathematical problem to be solved by the Schrödinger equation in cylindrical coordinates of a charged quantum particle confined to the symmetry axis of an electrically charged ring. Using the variational method, we determine the lowest energy level due to the potential generated by the charged ring. In Section IV, by using perturbation theory, we calculate the first eigenstates related to the interaction between an electron on the symmetry axis of an electrically charged ring. In Section V, we present the numerical results obtained by using the Numerov method. In this section we present some spectrum levels. Finally, in Section VI, we present the conclusion and perspectives.

%%%%%%%%%%%%%%%%%%%%%%%%%%%%%%%%%%%%%%%%%%%
%%%%%%%%%%%%%%%%%%%%%%%%%%%%%%%%%%%%%%%%%%%
%%%%%%%%%%%%%%%%%%%%%%%%%%%%%%%%%%%%%%%%%%%
%%%%%%%%%%%%%%%%%%%%%%%%%%%%%%%%%%%%%%%%%%%
%%%%%%%%%%%%%%%%%%%%%%%%%%%%%%%%%%%%%%%%%%%
%%%%%%%%%%%%%%%%%%%%%%%%%%%%%%%%%%%%%%%%%%%

\section{The Quantum Corral Problem}
%\indent

In this section, we present a theoretical quantum description about the quantum nanostructure called quantum corral. We address this system because our problem, an electrically particle charged on the symmetry axis of a electrically charged quantum ring, is a version more complicated of the quantum corral. The results presented in this section will be used further.

\begin{figure}[!ht]
%\begin{minipage}[!h]{0.4\linewidth}
\includegraphics[width=0.6\linewidth]{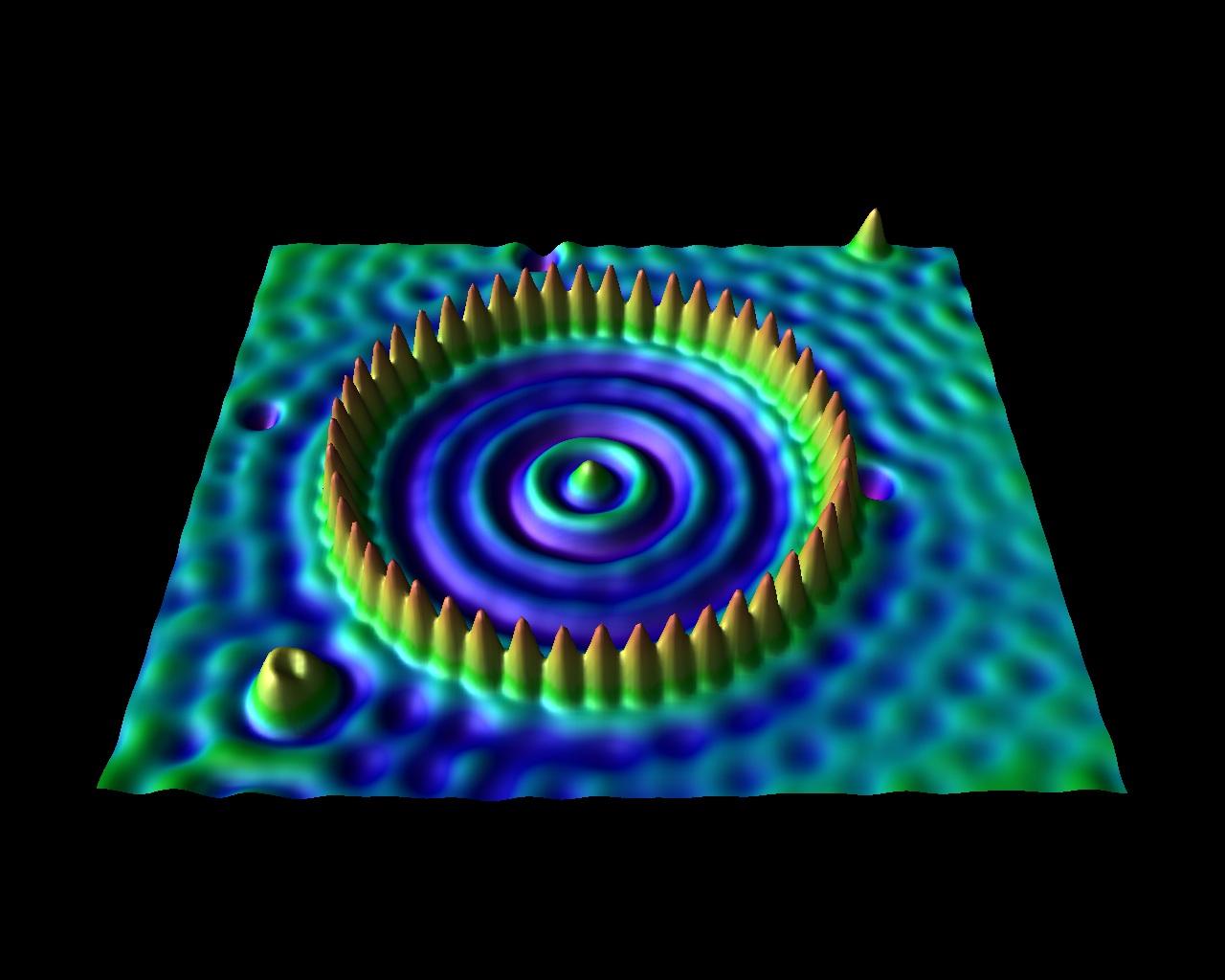}%\vspace{1.0cm}
\caption{Quantum Corral - Image obtained by the Scanning Tunneling Microscope available in the nisenet.org \cite{nisenet}. The radius of the corral is 7,1 nm.}
\label{curral_quantico}
%\end{minipage}
\end{figure}

The hypothetical nanostructure that we treat in this work is similar to ``quantum corral'' originally proposed by Eigler and collaborators in 1993 by manipulation of individual atoms using electron microscopy, specifically using a technique called Scanning Tunneling Microscopy (STM). In the model studied by Donald M. Eigler's, quantum corral was formed from 48 iron atoms in a circle on a copper surface under very high vacuum and very low temperature conditions, around 4 K. This experiment proved to be a brilliant example of nano-objects formation by manipulating individual atoms by STM. As it can be noted in Figure \ref{curral_quantico}, quantum corral works as a two-dimensional cylindrical quantum well that confines the surface electronic state. The circular waves observed inside the corral in the STM image are standing waves of electron density, whose existence is predicted by the Schrödinger equation for these specific boundary conditions \cite{Eigler1, Eigler2, Eigler3}.

The mathematical solution of the Schrödinger equation of the quantum nanostructure of the quantum corral is obtained in two dimensions using polar coordinates. The analysis contained in this section is closer to the solution presented by Donald M. Eigler and colleagues in 1993 \cite{Eigler1, Eigler3}. In \cite{Eigler3}, the authors propose that the quantum corral, seen in Figure \ref{curral_quantico}, is treated as an impassable circular rigid barrier, similar to the problem of the particle in a well of infinite potential. An electron trapped inside the inner region of the quantum corral will be described by a two-dimensional wave function in polar coordinates. As with the problem of the particle in a well of infinite potential, inside the corral the particle is free, that is, the potential $V$ is zero. The 2D Schrödinger equation for a particle confined in the quantum corral in polar coordinates is described as
\begin{equation}
\label{Schrodinger_2D}
 -\frac{\hbar^2}{2m}\left[\frac{1}{\rho}\frac{\partial}{\partial \rho}\left(\rho\frac{\partial }{\partial \rho} \right) + \frac{1}{\rho^2} \frac{\partial^2}{\partial \phi^2} \right]\psi(\rho,\phi) = E\psi(\rho,\phi).
\end{equation}
The differential operator is the kinetic energy term of the Hamiltonian operator in two-dimensional polar coordinates given in terms of the momentum operator, $\bm{p} = -i\hbar \nabla$, where the squared module is
\begin{equation}
 |\bm{p}|^2 = -\hbar^2 \nabla \cdot \nabla. \nonumber
\end{equation}
In polar coordinates the total momentum related to the particle trapped in the quantum corral is given by
\begin{equation}
\label{momento_curral_quantico}
 |\bm{p}|^2 = p_{\rho}^2 + \frac{L_{z}^2}{\rho^2}.
\end{equation}

The term $L_z$ is the angular momentum operator, which is perpendicular to the polar plane \cite{Goswami}. In Cartesian coordinates the angular momentum is given by $L_z = xp_{y}-yp_{x}$. Transforming from Cartesian to polar coordinates it leads to
\begin{equation}
 L_{z} = -i\hbar \frac{\partial}{\partial \phi}.
\end{equation}
The other component of the moment in the equation (\ref{momento_curral_quantico}) is given by $p_{\rho} = -i\hbar\dfrac{\partial}{\partial \rho}$.
Then, we have
\begin{equation}
 \bm{p} = -i\hbar \left(\hat{\bm \rho} \dfrac{\partial}{\partial \rho} + \hat{\bm \phi}\frac{1}{\rho}\frac{\partial}{\partial \phi}\right) = -i\hbar\nabla.
\end{equation}
The magnitude of the moment squared, $|\bm{p}|^2 $, with these differential terms, results in the differential operator of the Schrödinger equation (\ref{Schrodinger_2D}).

To solve  equation (\ref{Schrodinger_2D}), we use the method of separation of variables, denoting the wave function as $\psi(\rho,\phi) = P(\rho)\Phi(\phi)$. Substituting this expression in equation (\ref{Schrodinger_2D}) we get
\begin{eqnarray}
 \label{Schrodinger_2D_1}
 \frac{1}{P}\left(\frac{d^2 P}{d\rho^2} + \frac{1}{\rho} \frac{d P}{d\rho}\right) + \frac{1}{\rho^2\Phi}\frac{d^2\Phi}{d\phi^2} = -k^2,
\end{eqnarray}
where $k^2 = \dfrac{2mE}{\hbar^2}$. We rewrite the above equation as
\begin{equation}
\label{Schrodinger_2D_2}
 -\frac{1}{P}\left[\rho^2\frac{d^2 P}{d\rho^2} + \rho \frac{d P}{d\rho}\right] -k^2\rho^2 =  \frac{1}{\Phi}\frac{d^2\Phi}{d\phi^2} = - \nu^2.
\end{equation}
The right side of the above equation results in the well-known solution
\begin{equation}
 \label{funcao_momento_angular_1}
 \Phi = \Phi_{0}e^{i\nu\phi},
\end{equation}
where $\nu= 0,\pm 1,\pm 2,\cdots$. The above function is the solution for the quantum orbital angular momentum of an electron in the quantum corral, where $\nu$ is the integer number of the quantization of the angular momentum.

\begin{figure}[!ht]
%\begin{minipage}[!h]{1.0\linewidth}
\includegraphics[width=1.0\linewidth]{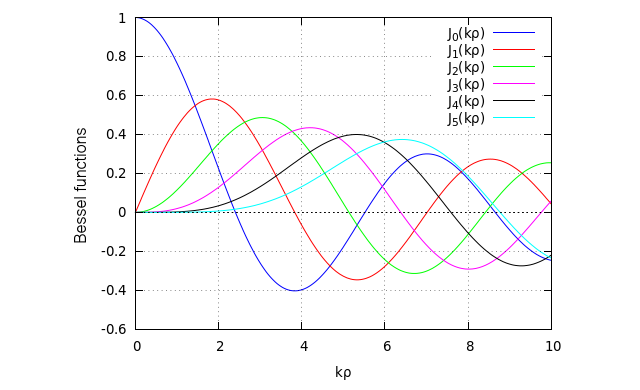}%\vspace{1.0cm}
\caption{First kind Bessel functions for $\nu =0,1,\cdots, 5$.}
\label{funcoes_bessel}
%\end{minipage}
\end{figure}

In this sense, other differential equation that we must solve from the equation
(\ref{Schrodinger_2D_2}) is the Bessel equation \cite{Arfken} given by
\begin{equation}
\label{equacao_Bessel}
 \rho^2\frac{d^2 P}{d\rho^2} + \rho \frac{d P}{d\rho} +(k^2\rho^2 - \nu^2)P = 0.
\end{equation}
The solution of this differential equation above is the Bessel functions,
\begin{equation}
\label{solucao_radial}
 P(\rho) = C_{1} J_{\nu}(k\rho) + C_{2} Y_{\nu}(k\rho) .
\end{equation}
The Bessel functions of the second kind $Y_{\nu}(k\rho)$ are singular in origin. Then, the divergence is avoid by choice $C_{2}=0$.  The two-dimensional wave function for a particle confined in the quantum corral will then be given by multiplying the solution presented in equation (\ref{funcao_momento_angular_1}) and the first-kind Bessel functions $J_{\nu}(k\rho)$
\begin{equation}
\label{funcao_onda_curral_quantico}
 \psi(\rho,\phi) = A  J_{\nu}(k\rho)e^{i\nu\phi},
\end{equation}
where the factor $A$ is a constant to be determined by normalizing the wave function.
The Bessel functions describing radial solutions are presented in Figure \ref{funcoes_bessel}.

Similar to the problem of a particle trapped in an infinite potential, the zeros of the Bessel functions quantize the energy of the quantum corral system. Looking at Figure \ref{curral_quantico}, we must impose for a particle trapped in the quantum corral that the wave function must be null in the impassable walls of the quantum corral. If the radius of the ring is $\rho=R$ the wavefunction $\psi$  must be zero in this region, i.e.
\begin{equation}
 \psi(R,\phi) = A  J_{\nu}(kR)e^{i\nu\phi} = 0.
\end{equation}
Initially, we choose the quantized angular momentum equal to zero. So for energy levels with $\nu=0$ we have that
\begin{equation}
 J_{0}(kR)= 0.
\end{equation}
The first point where the Bessel function $J_{0}$ cancels, see the graphic in Figure \ref{funcoes_bessel}, is $kR \approx 2,40483$. According to the definition  $k^2 = \dfrac{2mE}{\hbar^2}$, this results that the fundamental energy level is calculated as
\begin{equation}
 \frac{2mER^2}{\hbar^2} = 5,78321 \hspace*{0.5cm} \mbox{or} \hspace*{0.5cm} E = 2,89160 \frac{\hbar^2}{mR^2}.\nonumber
\end{equation}
The next zeros of the Bessel function $J_{0}(k\rho)$ are:
5,52008; 8,65373; 11,79153 and so on. For every zero in the function
$J_{0}$, it results in a quantized energy level of the quantum corral system. We can then index the energy in terms of the quantum numbers $p$ related to the zeros of the Bessel functions and $\nu$ related to the quantized angular momentum, it leads to
\begin{equation}
\label{energia_curral_quantico}
 E_{p\nu} = \frac{\hbar^2}{2m} k^2_{p\nu}.
\end{equation}
The first four values of $k^2_{p\nu}$ ($p=1,2,3,4$) that quantize the energy in the quantum corral system with zero angular momentum ($\nu=0$) are those ones shown in the sequence from the zeros of the Bessel function $J_{0}(k\rho)$,
\begin{eqnarray}
\label{valores_de_k}
 k_{10} = \frac{2,40483}{R}; \hspace*{0.5cm} k_{20} = \frac{5,52008}{R};\cr\cr
 k_{30} =\frac{8,65373}{R}; \hspace*{0.5cm}k_{40} =\frac{11,79153}{R}.
\end{eqnarray}
If it is necessary to calculate the energy values for non-zero quantum angular momentum ($\nu \neq 0$), it is possible to query values with the zeros of the tabled Bessel functions. In references \cite{Arfken,Wolfram} the first six zero values of the functions are displayed $J_{0}$ to $J_{5}$.

We observe in Figure \ref{curral_quantico} the existence of concentric waves internal to the quantum corral that corresponds to a possible quantum state given by the probability density obtained from a solution of the equation (\ref{funcao_onda_curral_quantico}).
In the references \cite{Eigler1,Eigler3}, the authors concluded that the image of concentric waves in Figure \ref{curral_quantico} is an electronic state surface of known effective mass measured in the experiment.

%%%%%%%%%%%%%%%%%%%%%%%%%%%%%%%%%%%%%%%%%%%
%%%%%%%%%%%%%%%%%%%%%%%%%%%%%%%%%%%%%%%%%%%
%%%%%%%%%%%%%%%%%%%%%%%%%%%%%%%%%%%%%%%%%%%
%%%%%%%%%%%%%%%%%%%%%%%%%%%%%%%%%%%%%%%%%%%
%%%%%%%%%%%%%%%%%%%%%%%%%%%%%%%%%%%%%%%%%%%
%%%%%%%%%%%%%%%%%%%%%%%%%%%%%%%%%%%%%%%%%%%

\section{Charged quantum particle on the axis of symmetry of an charged ring}

\indent

In this section we present the problem of a hypothetical nanostructure similar to the quantum corral. This structure must be an electrically charged circular nano-object. In macroscopic terms, it is an electrically charged ring very well discussed in classical electrodynamics \cite{Jackson}. Unlike the quantum corral discussed in the previous section, where the atoms that form the corral are on a metallic surface, the charged ring must have an empty circular interior, free so that the electronic states confined to the inner circular region also have a perpendicular degree of freedom to the circular area of the ring.
%%%%%%%%%%%%%%%%%%%
In principle we will state that the potential energy in cylindrical coordinates $(\rho,\phi,z)$ is given by
$V(\rho,\phi,z)$ with following boundary conditions,
\begin{equation}
 \label{boundary_conditions}
 V = \begin{cases}
      +\infty \hspace*{1cm} \mbox{for}~\rho = R~\mbox{and}~z \approx 0;\cr
      V(z) \hspace*{1cm} \mbox{for}~\rho<R ~\mbox{and small values for}~z.
     \end{cases}
\end{equation}
These boundary conditions state that: (i) the potential energy does not depend on the coordinate $\phi$; (ii) similarly to the quantum corral problem, at position $z=0$ and $\rho=R$ (radius ring), the potential function $V(R,\phi,0)$ is an infinite potential barrier, that is, the atoms that make up the ring act as a containment for the particle. In this case, the wave function of the particle confined in the inner region to the area of the ring is a stationary state symmetrically distributed in turn of the axis $z$; (iii) considering small oscillations on the axis of symmetry in the region $\rho<R$ and small values for coordinate $z$, the potential is given by the electrostatic potential produced by a ring charged with an electric charge $q$.

Let us analyze the problem of a charged quantum particle, such as an electron on the symmetry axis $z$ perpendicular to the circular area of the ring. In this way, the electrical potential is given by
\begin{equation}
 \label{potential_ring_3}
{\cal V}(z) = \frac{q}{4\pi\epsilon_0}\frac{1}{(z^2+ R^2)^{1/2}}.
\end{equation}
This simplified solution is very well addressed in the basic  electromagnetism \cite{Sears}.
From this electrostatic potential of a charged ring, we develop the quantum problem of a confined charged particle.
In similar way of the quantum corral problem, we assume that the particle is confined to the inner circular region, so that the electronic states are amplitudes of probability concentric to the axis $z$.
Let us apply this quantum energetic potential to the Schrödinger equation in cylindrical coordinates $(\rho,\phi,z)$,
\begin{equation}
\label{Schrodinger_1}
 -\frac{\hbar^2}{2m}\nabla^2\psi(\rho,\phi,z) +V(z)\psi(\rho,\phi,z) = E\psi(\rho,\phi,z).
\end{equation}
In cylindrical coordinates, the Laplacian differential operator is given by
\begin{equation}
\label{laplaciano}
 \nabla^2\psi = \frac{1}{\rho}\frac{\partial}{\partial \rho}\left(\rho\frac{\partial \psi}{\partial \rho} \right) + \frac{1}{\rho^2} \frac{\partial^2\psi}{\partial \phi^2}+\frac{\partial^2\psi}{\partial z^2} .
\end{equation}
Using the method of separating variables by
\begin{equation}
 \psi(\rho,\phi,z) = P(\rho)\Phi(\phi){\cal Z}(z), \nonumber
\end{equation}
and using Schrödinger equation (\ref{Schrodinger_1}), then, we obtain
\begin{eqnarray}
 \label{Schrodinger_2}
 \frac{1}{P}\left(\frac{d^2 P}{d\rho^2} + \frac{1}{\rho} \frac{d P}{d\rho}\right) + \frac{1}{\rho^2\Phi}\frac{d^2\Phi}{d\phi^2} = \cr
 -\frac{1}{\cal Z}\frac{d^2 {\cal Z}}{dz^2} + \frac{2m}{\hbar^2}\left[ V(z) - E\right] = -k^2.
\end{eqnarray}
The first part of this equation is exactly the same as equation (\ref{Schrodinger_2D_1}) of the quantum corral problem discussed in the previous section. The solution to the azimutal angular variable function is the function $\Phi$ obtained in equation (\ref{funcao_momento_angular_1}), so, that is $\Phi= \Phi_0 e^{i\nu\phi}$, where $\Phi_0$ is a constant and $\nu= 0,\pm 1,\pm 2,\cdots$.
The differential equation for the function $P$ is the same Bessel differential equation (\ref{equacao_Bessel}), whose solution to the problem must be the first kind Bessel functions $P = J_{\nu}(k\rho)$.

\begin{figure}[ht]
%\begin{minipage}[!h]{0.4\linewidth}
\includegraphics[width=1.0\linewidth]{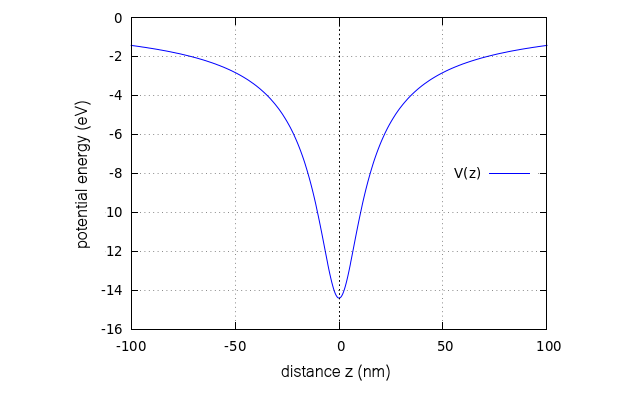}%\vspace{1.0cm}
\caption{Potential energy between the electrically charged ring and an electron on the axis of symmetry. In the graph, the electric charge of the ring is $q=+100 e$ and radius $R =10$ nm. }
\label{energia_Vz}
%\end{minipage}
\end{figure}

The equation for function $Z(z)$ is given by second part of equation (\ref{Schrodinger_2}),
\begin{equation}
\label{Schrodinger_3}
 -\frac{\hbar^2}{2m}\frac{d^2 {\cal Z}(z)}{dz^2} +V(z){\cal Z}(z) = \left(E - \frac{\hbar^2 k^2}{2m}\right) {\cal Z}(z).
\end{equation}
The factor $k$ in last equation is the same $k_{p\nu}$ discussed in the previous section of the quantum corral problem, due
$J_{\nu}(k_{p\nu}R) =0$. Then, factor $k_{p\nu}$ determines the possible eigenvalues for the particle confined in the region inside the ring of radius $R$.
Lets define
\begin{equation}
\label{energia_anel_1}
 E_{n} = E - \frac{\hbar^2 k^2_{p\nu}}{2m}
\end{equation}
Note that the second term of the above equation is the quantized levels of energy of the quantum corral obtained in equation (\ref{energia_curral_quantico}),  $E_{p\nu} =  \dfrac{\hbar^2 k^2_{p\nu}}{2m}$. Therefore, these results are the same possible energy levels for the quantum particle confined in the inner region of the radius ring $R$. Then we rewrite the equation (\ref{Schrodinger_3}) as
\begin{equation}
\label{Schrodinger_4}
 -\frac{\hbar^2}{2m}\frac{d^2 {\cal Z}(z)}{dz^2} +V(z){\cal Z}(z) = E_{n} {\cal Z}(z).
\end{equation}
The above equation is a usual one-dimensional Schrödinger equation.
We can replace the energy potential (\ref{potential_ring_3}) in the Schrödinger equation (\ref{Schrodinger_4}). In this work we assume that the charge of the ring is $+q$ and that the particle is an electron with a negative elementary charge $-e$. The potential energy is given
\begin{equation}
\label{energia_potencial}
  V(z) = - V_0\left( 1+ \frac{z^2}{R^2}\right)^{-1/2},
\end{equation}
where
\begin{equation}
\label{energia_V0}
 V_0= \frac{eq}{4\pi\epsilon_0 R}.
\end{equation}
In Figure \ref{energia_Vz}, we see the graph of potential energy $V(z)$.

Finally we rewrite the Schrödinger equation (\ref{Schrodinger_4}) as
\begin{equation}
 \label{Schrodinger_5}
 -\frac{\hbar^2}{2m}\frac{d^2 {\cal Z}(z)}{dz^2} -V_0\left( 1+ \frac{z^2}{R^2}\right)^{-1/2} {\cal Z}(z) = E_{n} {\cal Z}(z).
\end{equation}
Solving the Schrödinger equation (\ref{Schrodinger_5}), we obtain the possible solutions related to the degree of freedom of the variable $z$: the eigenfunctions $Z(z)$ and the eigenvalues $E_{n}$. These results must be recombined with the already known radial and angular solutions, which are the same as for the quantum corral problem. The energy eigenvalues $E_{n}$ after being calculated are used in equation (\ref{energia_anel_1}) to determine the energy levels for the system considered.  
%It is important to note that the boundary conditions established in this section for $\psi(\rho,\phi, z)$ are not general. This specific choice is due the physical interpretation is easier.

%%%%%%%%%%%%%%%%%%%%%%%%%%%%%%%%%%%%%%%%%%%
%%%%%%%%%%%%%%%%%%%%%%%%%%%%%%%%%%%%%%%%%%%
%%%%%%%%%%%%%%%%%%%%%%%%%%%%%%%%%%%%%%%%%%%
%%%%%%%%%%%%%%%%%%%%%%%%%%%%%%%%%%%%%%%%%%%
%%%%%%%%%%%%%%%%%%%%%%%%%%%%%%%%%%%%%%%%%%%

\section{Perturbation Method}
\indent

The perturbation theory method of the quantum mechanics is an important tool to describe theoretical and real quantum systems when the hamiltonian operator of the quantum system has a degree of complexity that makes it difficult to analytically solve the Schrödinger equation. In fact, few Quantum Mechanics problems have exact solutions, such as the hydrogen atom, the quantum harmonic oscillator and the particle in a box. Perturbation Theory starts from the knowledge of a known exact solution to generate approximate analytical solutions of a more complicated quantum problem \cite{Griffths,Cohen}. Using the method of the Theory of Perturbation, we can separate in the Hamiltonian of the problem, a part that corresponds to an exact solution and other terms that correspond to the perturbative terms. For our problem, from the quantum particle (electron) interacting with the circular nano-object, the electrically charged ring, we can expand the potential energy function into a power series. Substituting this result in equation (\ref{potential_ring_3}) to obtain a subdivision or sector of the hamiltonian of the simple harmonic oscillator, proportional to the term $z^2$, which corresponds to the exact solution and a term proportional to $z^4$ which corresponds to the perturbative term. As previously stated, for small distances $z$ between the particle and the ring, $z<R$, this first perturbative term will make it possible to calculate with the Perturbation Theory the corrections of the eigenvalues of energies and of the wave functions for the first levels of energy corresponding to the Schrödinger equation (\ref{Schrodinger_5}).
The potential energy given in equation (\ref{energia_potencial}) is given by
\begin{equation}
 V(z) = -V_0\left( 1+ \frac{z^2}{R^2}\right)^{-1/2},\nonumber
\end{equation}
where the first terms of the series expansion of the potential energy above are given by
\begin{equation}
 V(z) = -V_0\left( 1- \frac{z^2}{2R^2} + \frac{3z^4}{8R^4} -\frac{15z^6}{48 R^6}+\cdots \right).
\end{equation}
The analysis with the Perturbation Theory with this potential energy will be restricted to the order of $z^6$, so we rewrite the Schrödinger equation (\ref{Schrodinger_5}) as,
\begin{eqnarray}
 \label{Schrodinger_6}
 \left[-\frac{\hbar^2}{2m}\frac{d^2 }{dz^2} +\frac{V_0}{2R^2} z^2\right]{\cal Z}- \left[\frac{3V_0}{8R^4}z^4 - \frac{15 V_0}{48 R^6}z^6\right]{\cal Z} \cr = (E_{n}+ V_0) {\cal Z}.
\end{eqnarray}
Note that the first term in square brackets is similar to the exact solution Hamiltonian of the simple harmonic oscillator. Let us organize the above equation by making the following definition,
\begin{equation}
 \frac{V_0}{R^2} = m\omega^2,
\end{equation}
where we explain the angular velocity of the harmonic oscillations: $\omega^2 = \dfrac{V_0}{mR^2}$. The perturbation factor proportional to $z^4$ has the constant simplified through the definition,
\begin{equation}
 \frac{3V_0}{8R^4}=\Lambda_1,
\end{equation}
and the perturbative factor proportional to $z^6$ has the constant simplified through the definition,
\begin{equation}
 \frac{15V_0}{48R^6}=\Lambda_2.
\end{equation}
We also simplify,
\begin{equation}
\label{energia_anel_2}
 E_{n}+V_0 = {\cal E}.
\end{equation}
Then we can write the Schrödinger equation (\ref{Schrodinger_6}) as follows
\begin{eqnarray}
  \label{Schrodinger_7}
 \left[-\frac{\hbar^2}{2m}\frac{d^2 }{dz^2} +\frac{m\omega^2}{2} z^2\right]{\cal Z}(z)- \left[\Lambda_1 z^4 - \Lambda_2 z^6 \right]{\cal Z}(z) \cr = {\cal E} {\cal Z}(z).
\end{eqnarray}
In this equation we can visualize the Hamiltonian of the simple harmonic oscillator adding to the perturbation term $W=-\Lambda_1 z^4 + \Lambda_2 z^6$.
From here we can use the Perturbation Theory method to calculate some of the first eigenstates.

The Schrödinger equation of the simple harmonic oscillator is
\begin{equation}
\left[-\frac{\hbar^2}{2m}\frac{d^2 }{dz^2} +\frac{m\omega^2}{2} z^2\right]{\cal Z}(z) =\hbar\omega \left(n+\frac{1}{2} \right){\cal Z}(z),\nonumber
\end{equation}
and has a well determined exact solution, with quantized energies given by ${\cal E}^{(0)}_{n} = \hbar\omega \left(n+\frac{1}{2} \right) $. Then we can analytically solve the problem of the Schrödinger equation with partubation theory given in equation (\ref{Schrodinger_7}).
To solve the perturbative problem, we must start the analysis using the creation and destruction operators given by \cite{Cohen}:
\begin{equation}
\label{operadores_1}
a^{\dag} = \sqrt{\frac{m\omega}{2\hbar}}\left(z-\frac{ip_z}{m\omega}\right), \hspace*{0.5cm} a = \sqrt{\frac{m\omega}{2\hbar}}\left(z + \frac{ip_z}{m\omega}\right),
\end{equation}
where we can define from these operators above the operator $N$,
\begin{equation}
\label{operadores_2}
a^{\dag} a = N,
\end{equation}
with $N | n \rangle  = n| n \rangle $, where $n$ is an eigenvalue of the operator $N$.
The algebra of these operators are visualized below,
\begin{equation}
\label{operadores_3}
[a,a^{\dag} ] = 1,~~[N,a^{\dag} ] = a^{\dag},~~\mbox{and} ~~ [N,a] = -a.
\end{equation}
%\begin{equation}
%\label{operadores_4}
%[N,a^{\dag} ] = a^{\dag},
%\end{equation}
%\begin{equation}
%\label{operadores_5}
%[N,a] = -a.
%\end{equation}
We also add the operations of the creation and destruction operators respectively in the eigenvectors $| n \rangle$,
\begin{equation}
\label{operadores_6}
a^{\dag}\, | n \rangle = \sqrt{n+1} \,| n +1 \rangle \hspace*{0.5 cm} \mbox{and} \hspace*{0.5 cm} a\,| n \rangle = \sqrt{n} \,| n -1 \rangle.
\end{equation}
%\begin{equation}
%\label{operadores_7}
%a^{\dag}\, | n \rangle = \sqrt{n+1} \,| n +1 \rangle
%\end{equation}
From the operators given in equation (\ref{operadores_1}) we can express the $z$ operator in the form
\begin{equation}
z = \sqrt{\frac{\hbar}{2m\omega}}\left(a+ a^{\dag} \right).
\end{equation}
With this result, we can calculate the perturbative term $W=-\Lambda_1 z^4 + \Lambda_2 z^6$ in terms of the operator above. We calculate $z^4$ which results in
\begin{eqnarray}
\label{z_quarta}
 z^4  &=& \left(\frac{\hbar}{2m\omega}\right)^2 (a^4+ a^{\dag\, 4} + 6 a^2 - 2a^{\dag\, 2}  +  4Na^{\dag\, 2} \cr
 & & + 4N a^2 +6N^2+ 6N+3 ),
\end{eqnarray}
and the term $z^6$ which results in
\begin{eqnarray}
\label{z_sexta}
z^6  &=& \left(\frac{\hbar}{2m\omega}\right)^3 [a^6+ a^{\dag\, 6} + (6N+15)a^4 + (6N-9) a^{\dag\, 4} \cr
& & + (15N^2+45N+45)a^2 + (15N^2 - 15N+ 15)a^{\dag\, 2} \cr
 & & + 20N^3+30N^2+40N+15].
\end{eqnarray}

According to perturbation theory, the energy eigenvalues for the Schrödinger equation (\ref{Schrodinger_7}) is given by
\begin{equation}
\label{energia_pertubacao_01}
 {\cal E}_{n}  =  {\cal E}^{(0)}_{n} + {\cal E}^{(1)}_{n} + {\cal E}^{(2)}_{n} + \cdots
\end{equation}
The first term on the right side of the equation refers to the energy levels of the simple harmonic oscillator (exact solution), where ${\cal E}^{(0)}_{n} = \hbar\omega \left(n+\frac{1}{2} \right) $. The other terms on the right side, $ {\cal E}^{(1)}_{n} + {\cal E}^{(2)}_{n}$, refer to the first and second order corrections respectively, which are calculated with the Perturbation Theory. For the corrections in the energy levels up to the first order we must calculate
\begin{equation}
{\cal E}^{(1)} = \langle n|\, W \, | n \rangle = - \Lambda_1 \langle n|\,z^4  \, | n \rangle + \Lambda_2 \langle n|\,z^6  \, | n \rangle.
\end{equation}
Considering the $z^4$ operator in the equation (\ref{z_quarta}) and $z^6$ of the equation (\ref{z_sexta}) in the energy calculation ${\cal E}^{(1)}$, the non-zero values are given by
\begin{eqnarray}
{\cal E}^{(1)}_{n} &=& -\frac{3}{32} \frac{(\hbar\omega)^2}{V_0}(6n^2+6n +3) \cr
& & + \frac{15}{384} \frac{(\hbar\omega)^3}{V_0^2}(20n^3 + 30n^2 + 40n +15).
\end{eqnarray}
It is important to note that in the above equation, the first-order correction on the $z^6$ term is proportional to $\frac{(\hbar\omega)^3}{V_0^2}$.
As we will see below, corrections in the second-order energy levels of the perturbation $\Lambda_1 z^4$ are also multiples of $\frac{(\hbar\omega)^3}{V_0^2}$. The second-order corrections in the $z^4$ operator are relevant and we must perform the calculations by using equation below,
\begin{equation}
{\cal E}^{(2)}_{n} = \sum_{n\neq n'} \frac{|\,\langle n|\, W \, | n' \rangle\,|^2}{{\cal E}_{n'}-{\cal E}_{n}}.
\end{equation}
The terms of perturbation
$\langle n|\, W \, | n' \rangle = \Lambda\langle n|\, z^4 \, | n' \rangle$ that contribute to second-order corrections are off-diagonal terms. From the equation (\ref{z_quarta}) we highlight that the terms that contribute to these calculations are the terms $(a^4+ a^{\dag\, 4} + 6 a^2 - 2a^{\dag\, 2}  +  4Na^{\dag\, 2} + 4N a^2) $ so that the possible results for $\langle n|\, W \, | n' \rangle$ are
\begin{small}
 \begin{equation}
 \langle n-2| W| n \rangle =  - \frac{3(\hbar\omega)^2}{32 V_0} \left[ (4n-2) \sqrt{n(n-1)} \right], \nonumber
\end{equation}
 \begin{equation}
 \langle n+2| W | n \rangle = - \frac{3(\hbar\omega)^2}{32 V_0}  \left[  (4n+6) \sqrt{(n+1)(n+2)} \right],\nonumber
\end{equation}
 \begin{equation}
 \langle n-4| W | n \rangle = -\frac{3(\hbar\omega)^2}{32 V_0} \left[ \sqrt{n(n-1)(n-2)(n-3)}  \right],\nonumber
\end{equation}
 \begin{equation}
 \langle n+4| W | n \rangle = - \frac{3(\hbar\omega)^2}{32 V_0}  \left[ \sqrt{(n+1)(n+2)(n+3)(n+4)}  \right].\nonumber
\end{equation}
\end{small}
\\
With the above equations we can determine the second order corrections for some of the first energy levels. For the ground state, $n=0$ , we get two of these above non-zero terms,
\begin{equation}
\label{2W0}
 \langle 2| W | 0 \rangle =  - \frac{3(\hbar\omega)^2}{32 V_0} (6\sqrt{2})
\end{equation}
and
\begin{equation}
\label{4W2}
 \langle 4| W | 0 \rangle = - \frac{3(\hbar\omega)^2}{32 V_0} (2\sqrt{6}),
\end{equation}
so that,
\begin{eqnarray}
 {\cal E}^{(2)}_0 &=& \frac{| \langle 2| W| 0 \rangle|^2 }{{\cal E}_0 - {\cal E} _2} +  \frac{| \langle 4| W | 0 \rangle| ^2}{{\cal E}_0 - {\cal E} _4}\cr
 &=& \frac{9}{1024}\frac{(\hbar\omega)^4}{V_0^2}\left(\frac{72}{(-2)\hbar\omega} + \frac{24}{(-4)\hbar\omega}\right),\nonumber
\end{eqnarray}
which numerically results in
\begin{equation}
\label{E_fundamental_2}
 {\cal E}^{(2)}_0 = -0,369140625 \frac{(\hbar\omega)^3}{V_0^2}.
\end{equation}
For the first excited state, $n=1$,  we obtain,
\begin{equation}
\label{3W1}
 \langle 3| W | 1 \rangle =  - \frac{3(\hbar\omega)^2}{32 V_0} (10\sqrt{6})
\end{equation}
and
\begin{equation}
\label{5W1}
 \langle 5| W | 1 \rangle = - \frac{3(\hbar\omega)^2}{32 V_0} (\sqrt{120}),
\end{equation}
so that,
\begin{eqnarray}
 {\cal E}^{(2)}_1 &=& \frac{| \langle 3| W| 1 \rangle|^2 }{{\cal E}_1 - {\cal E} _3} +  \frac{| \langle 5| W | 1 \rangle| ^2}{{\cal E}_1 - {\cal E} _5}\cr
 &=& \frac{9}{1024}\frac{(\hbar\omega)^4}{V_0^2}\left(\frac{600}{(-2)\hbar\omega} + \frac{120}{(-4)\hbar\omega}\right)\nonumber
\end{eqnarray}
which numerically results in
\begin{equation}
\label{E_excitado_1}
 {\cal E}^{(2)}_1 = -2,900390625 \frac{(\hbar\omega)^3}{V_0^2}.
\end{equation}

%%%%%%%%%%%%%%%%%%%%%%%%%%%%%%%%%%%%%%%%%%%%%%
%%%%%%%%%%%%%%%%%%%%%%%%%%%%%%%%%%%%%%%%%%%%%%
%%%%%%%%%%%%%%%%%%%%%%%%%%%%%%%%%%%%%%%%%%%%%%

For the second the excited state, $n=2$,  we obtain,
\begin{equation}
 \langle 0| W | 2 \rangle =  - \frac{3(\hbar\omega)^2}{32 V_0} (6\sqrt{2}) , \nonumber
\end{equation}
\begin{equation}
 \langle 4| W | 2 \rangle = - \frac{3(\hbar\omega)^2}{32 V_0} (14\sqrt{12}),  \nonumber
\end{equation}
and
\begin{equation}
 \langle 6| W | 2 \rangle = - \frac{3(\hbar\omega)^2}{32 V_0} (\sqrt{360}). \nonumber
\end{equation}
This leads to,
\begin{eqnarray}
 {\cal E}^{(2)}_2 &=& \frac{| \langle 0| W| 2 \rangle|^2 }{{\cal E}_2 - {\cal E} _0} +  \frac{| \langle 4| W | 2 \rangle| ^2}{{\cal E}_2 - {\cal E} _4} +  \frac{| \langle 6| W | 2 \rangle| ^2}{{\cal E}_2 - {\cal E} _6}\cr
 &=& \frac{9}{1024}\frac{(\hbar\omega)^3}{V_0^2}\left(\frac{72}{2}  + \frac{2352}{(-2)} + \frac{3600}{(-4)}\right), \nonumber
\end{eqnarray}
which numerically results in
\begin{equation}
\label{E_excitado_2}
 {\cal E}^{(2)}_2 = -10,81054688 \frac{(\hbar\omega)^3}{V_0^2}.
\end{equation}

%%%%%%%%%%%%%%%%%%%%%%%%%%%%%%%%%%%%%%%%%%%%%%
%%%%%%%%%%%%%%%%%%%%%%%%%%%%%%%%%%%%%%%%%%%%%%
%%%%%%%%%%%%%%%%%%%%%%%%%%%%%%%%%%%%%%%%%%%%%%

For the third the excited state, $n=3$, we obtain,
\begin{equation}
 \langle 1| W | 3 \rangle =  - \frac{3(\hbar\omega)^2}{32 V_0} (10\sqrt{6}), \nonumber
\end{equation}
\begin{equation}
 \langle 5| W | 3 \rangle = - \frac{3(\hbar\omega)^2}{32 V_0} (18\sqrt{20}),  \nonumber
\end{equation}
and
\begin{equation}
 \langle 7| W | 3 \rangle = - \frac{3(\hbar\omega)^2}{32 V_0} (\sqrt{840}),  \nonumber
\end{equation}
so that,
\begin{eqnarray}
 {\cal E}^{(2)}_3 &=& \frac{| \langle 1| W| 3 \rangle|^2 }{{\cal E}_3 - {\cal E} _1} +  \frac{| \langle 5| W | 3 \rangle| ^2}{{\cal E}_3 - {\cal E} _5} + \frac{| \langle 7| W | 3 \rangle| ^2}{{\cal E}_3 - {\cal E} _7}\cr
 &=& \frac{9}{1024}\frac{(\hbar\omega)^3}{V_0^2}\left(\frac{600}{2} + \frac{6480}{(-2)} +\frac{840}{(-4)} \right), \nonumber
\end{eqnarray}
which numerically is given by
\begin{equation}
\label{E_excitado_3}
 {\cal E}^{(2)}_3 = -27,68554688 \frac{(\hbar\omega)^3}{V_0^2}.
\end{equation}
By following this method we can move forward into other energy levels. We show the result of the second order correction for the next energy level, the fourth the excited state, $n=4$,
\begin{equation}
\label{E_excitado_4}
 {\cal E}^{(2)}_4 = -57,11132813 \frac{(\hbar\omega)^3}{V_0^2}.
\end{equation}

With these computations, we can calculate the quantized energies of the eigenstates of the charged particle interacting with the charged ring with corrections up to the second order as highlighted in the equation (\ref{energia_pertubacao_01}),
\begin{eqnarray}
 \label{energia_pertubacao_02}
 {\cal E}_{n}  & \approx & \hbar\omega \left(n+\frac{1}{2} \right)  -\frac{3}{32} \frac{(\hbar\omega)^2}{V_0}(6n^2+6n +3)\cr
 &+& \frac{15}{384} \frac{(\hbar\omega)^3}{V_0^2}(20n^3 + 30n^2 + 40n +15) + {\cal E}^{(2)}_{n}.
\end{eqnarray}
The terms for ${\cal E}^{(2)}_{n}$ ($n=0,1,\cdots 4$) were obtained in equations (\ref{E_fundamental_2})-(\ref{E_excitado_4}).
As stated earlier, let us analyze the results for an electron of charge $-e$ interacting with an electrically charged ring with charge $q$. By dimensioning the charged ring with the numerical values for the charge $q$ and the radius $R$, we will be able to  calculate numerically the energy levels through the equation above (\ref{energia_pertubacao_02}).
With the calculated values of ${\cal E}_{n}$, we can go back in the equation (\ref{energia_anel_2}) and calculate the resulting energy,
\begin{equation}
\label{energia_pertubacao_03}
 E_{n} = {\cal E}_{n} - V_{0}.
\end{equation}
And with the above values we return to the equation (\ref{energia_anel_1}) to finally obtain the energy spectrum of an electron interacting with the electrically charged ring, with the superposition of the quantum numbers $p$, $\nu$ and $n$,
\begin{equation}
\label{energia_anel_3}
 E_{p\nu n} = {\cal E}_{n}- \frac{\hbar^2 k^2_{p\nu}}{2m} - V_{0} .
\end{equation}

%%%%%%%%%%%%%%%%%%%%%%%%%%%%%%%%%%%%%%%%%%%%%%
%%%%%%%%%%%%%%%%%%%%%%%%%%%%%%%%%%%%%%%%%%%%%%
%%%%%%%%%%%%%%%%%%%%%%%%%%%%%%%%%%%%%%%%%%%%%%
%%%%%%%%%%%%%%%%%%%%%%%%%%%%%%%%%%%%%%%%%%%%%%
%%%%%%%%%%%%%%%%%%%%%%%%%%%%%%%%%%%%%%%%%%%%%%
%%%%%%%%%%%%%%%%%%%%%%%%%%%%%%%%%%%%%%%%%%%%%%
%%%%%%%%%%%%%%%%%%%%%%%%%%%%%%%%%%%%%%%%%%%%%%
%%%%%%%%%%%%%%%%%%%%%%%%%%%%%%%%%%%%%%%%%%%%%%
%%%%%%%%%%%%%%%%%%%%%%%%%%%%%%%%%%%%%%%%%%%%%%

To obtain the eigenfunctions of the Schrödinger equation (\ref{Schrodinger_7}) from the method of time-independent perturbation theory, the wave eigenfunctions ${\cal Z}(z)$ are obtained from the expansion
\begin{equation}
\label{autofuncoes_teoria_perturbacao}
 {\cal Z}_n(z) = \psi_n^{(0)}(z) + \psi_n^{(1)}(z) + \cdots
\end{equation}
where $\psi_n^{(0)}(z)$ are the eigenfunctions of the known exact solution. Our analysis for a charged electron interacting with the charged ring relies on the solution of the simple harmonic oscillator, then the unperturbed  wavefunctions $\psi_n^{(0)}(z)$ are the exact solutions of the simple harmonic oscillator:
\begin{equation}
\label{autofuncoes_OHS}
 \psi_n^{(0)}(\xi) = \left(\frac{m\omega}{\pi\hbar} \right)^{1/4}\frac{1}{\sqrt{2^n\, n!}}H_n(\xi)e^{-\xi^2/2},
\end{equation}
where
\begin{equation}
 \xi = \sqrt{\frac{m\omega}{\hbar}}\, z
\end{equation}
and $H_n(\xi)$ are polynomials of Hermite,
\begin{equation}
 \begin{matrix}
  H_0 = 1\cr
  H_1 = 2\xi \cr
  H_2 = 4\xi^2 -2 \cr
  H_3 = 8\xi^3 - 12\xi \cr
  H_4 = 16\xi^4 - 48\xi^2 +12 \cr
  H_5 = 32\xi^5 -160\xi^3 +120\xi.
 \end{matrix}
\end{equation}

The first order correction terms $\psi_n^{(1)}(x)$ in equation  (\ref{autofuncoes_teoria_perturbacao}) are calculated using the equation:
\begin{equation}
 \psi_n^{(1)}(x) = \sum_{n\neq k}\frac{\langle k |W| n \rangle}{{\cal E}_n^{(0)} - {\cal E}_k^{(0)}}\psi_k^{(0)}.
\end{equation}
The ground state wave function of the Schrödinger equation (\ref{Schrodinger_7}), corresponding to the electron interacting with the charged ring on the axis of symmetry $z$, will then be given by equation (\ref{autofuncoes_teoria_perturbacao}) with corrections to the first order.
The possible correction terms for the first order are given by,
\begin{equation}
 {\cal Z}_0 = \psi_0^{(0)} +  \frac{\langle 2 |W| 0 \rangle}{{\cal E}_0 - {\cal E}_2}\psi_2^{(0)} +  \frac{\langle 4 |W| 0 \rangle}{{\cal E}_0 - {\cal E}_4}\psi_4^{(0)}. \nonumber
\end{equation}
Using equation (\ref{2W0}) and  equation (\ref{4W2})   in the above equation, it follows that
\begin{eqnarray}
{\cal Z}_0
&=& \psi_0^{(0)} +  \frac{\frac{-3\,(\hbar\omega)^2}{32\,V_0}6\sqrt{2}}{(-2)\hbar\omega}\psi_2^{(0)} +  \frac{\frac{-3\,(\hbar\omega)^2}{32\,V_0}2\sqrt{6}}{(-4)\hbar\omega}\psi_4^{(0)}\cr
&=& \psi_0^{(0)} + 0,397747564\left(\frac{\hbar\omega}{V_0}\right)\psi_2^{(0)}\cr
&&+ 0,114819831 \left(\frac{\hbar\omega}{V_0}\right) \psi_4^{(0)}, \nonumber
\end{eqnarray}
then the resulting wave function for the ground state is
\begin{small}
\begin{eqnarray}
\label{autoestado_Z_0}
 {\cal Z}_0 &=& \left(\frac{m\omega}{\pi\hbar} \right)^{1/4}e^{-\xi^2/2}[1 + 0,140624999\left(\frac{\hbar\omega}{V_0}\right)(4\xi^2 - 2)\cr
 &+&  0,005859375\left(\frac{\hbar\omega}{V_0}\right) (16\xi^4 - 48\xi^2 +12)].
\end{eqnarray}
\end{small}

Let us calculate the wavefunction of the first perturbed excited eigenstate in first order, whose possible correction terms for the first order are shown in the equation below:
\begin{equation}
 {\cal Z}_1 = \psi_1^{(0)} +  \frac{\langle 3 |W| 1 \rangle}{{\cal E}_1 - {\cal E}_3}\psi_3^{(0)} +  \frac{\langle 5 |W| 1 \rangle}{{\cal E}_1 - {\cal E}_5}\psi_5^{(0)}. \nonumber
\end{equation}
Using the results of equation (\ref{3W1}) and equation (\ref{5W1}) in the above equation, we then get,
\begin{eqnarray}
{\cal Z}_1 &=& \psi_1^{(0)} +  \frac{\frac{-3\,(\hbar\omega)^2}{32\,V_0}10\sqrt{6}}{(-2)\hbar\omega}\psi_2^{(0)} +  \frac{\frac{-3\,(\hbar\omega)^2}{32\,V_0}2\sqrt{120}}{(-4)\hbar\omega}\psi_5^{(0)}\cr
&=& \psi_1^{(0)} + 1,148198317\left(\frac{\hbar\omega}{V_0}\right)\psi_3^{(0)}\cr
&+& 0,256744948 \left(\frac{\hbar\omega}{V_0}\right) \psi_5^{(0)}. \nonumber
\end{eqnarray}
Then the resulting wavefunction for the first excited state is given by,
\begin{small}
\begin{eqnarray}
\label{autoestado_Z_1}
 {\cal Z}_{1} &=& \left(\frac{m\omega}{\pi\hbar} \right)^{1/4}e^{-\xi^2/2}[1,414213562~\xi \cr  &+& 0,165728151\left(\frac{\hbar\omega}{V_0}\right)(8\xi^3 - 12\xi)\cr
 &+&  0,004143204\left(\frac{\hbar\omega}{V_0}\right) (32\xi^5 - 160\xi^3 +120\xi)].
\end{eqnarray}
\end{small}
Following this methodology we can obtain the other eigenstates for the Schrödinger equation (\ref{Schrodinger_7}) and with these eigenstates compose the complete solution to the problem of an electron interacting with an electrically charged ring. Thus, the complete solution to the wave function follows by,
\begin{equation}
 \psi_{p\nu n}(\rho,\phi,z) = A J_{\nu}(k_{p\nu}\rho) e^{i\nu\phi} Z_{n}(z), \nonumber
\end{equation}
where $p$ stands for the quantum number referring to the zeros of the Bessel functions, $\nu$ is the quantum number of the angular momentum and $n$ is the quantum number of the eigenstates of the Schrödinger equation (\ref{Schrodinger_7}), with two eigenstates calculated in the equations (\ref{autoestado_Z_0}) and (\ref{autoestado_Z_1}).

With the ring dimensions defined, $q=+100\,e$ and $R=10$ nm, using the preceding results, we obtain numerically the first five possible energy values $E_{n}$ (com $n=0,\cdots, 4$), in the table \ref{tabela2}, for the electron confined to the energetic potential of the charged ring.
\begin{table}[h]
%\begin{ruledtabular}
	\centering
	\caption{The first five energy levels of the eigenstates of the Schrödinger equation (\ref{Schrodinger_6}), by the method of the Theory of Perturbation.}
	\begin{tabular}{|c|c|} \hline \hline
	
	 n  & Energy  (eV) \\
	 \hline
	0 & -14,3478480690
 \\
	\hline
	1 & -14,2439826462
 \\
	\hline
	2 & -14,1410040363
 \\
	\hline
	3 & -14,0389321083
 \\
	\hline
	 4 & -13,9377867309
\\ \hline \hline
	\end{tabular}
		\label{tabela2}
	%\end{ruledtabular}	
	\end{table}

\begin{figure}[!ht]
%\begin{minipage}[!h]{0.4\linewidth}
\includegraphics[width=1.0\linewidth]{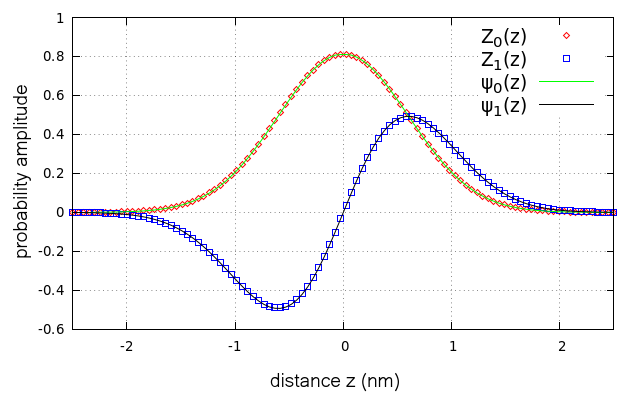}%\vspace{1.0cm}
\caption{Wave functions of the first two eigenstates. The electric charge of the ring is worth $q=+100~e$ and the radius $R =10$ nm.}
\label{autoestados_Z0_Z1_T_Perturbacao}
%\end{minipage}
\end{figure}

The graph in Figure \ref{autoestados_Z0_Z1_T_Perturbacao}  shows the first two eigenfunctions: $Z_{0}$ obtained in equation (\ref{autoestado_Z_0}) of the ground state and $Z_{1}$ obtained in the equation (\ref{autoestado_Z_1}) from the first excited state. We also plot the two unperturbed eigenfunctions of the quantum simple harmonic oscillator, $\psi_{0}$ and $\psi_{1}$. We observe that the two eigenfunctions $Z_{0}$ and $Z_{1}$ corrected to first order do not differ graphically from unperturbed functions $\psi_{0}$ and $\psi_{1}$. The first two eigenstates correspond to small oscillations that graphically are very close to the eigenstates of the simple harmonic oscillator.

The Perturbation Theory method is very useful to obtain the values of the eigenstates of a quantum problem. With the results obtained through the Perturbation Theory, we can use these results as guides in the more precise computational numerical methods that have the potential to refine the numerical values initially obtained through the Perturbation Theory.

%
%
%
%\newpage
\section{The Numerical Method of Numerov}

\indent

We will now use an efficient numerical method to solve certain types of second-order ordinary differential equations. The Numerical Method of Numerov \cite{Hairer} has proved to be very useful for numerically solving the Schrödinger equation \cite{Caruso}.
To solve the Schrödinger equation problem (\ref{Schrodinger_5}), with the values of the hypothetical example seen in the previous section, the Schrödinger equation to be solved is rewritten as:
\begin{equation}
\label{Schrodinger_8}
\frac{d^2 {\cal Z}(z)}{dz^2} + \left[\frac{2m}{\hbar^2}\frac{V_0}{ \sqrt{1+\frac{z^2}{R^2}}}+\frac{2m}{\hbar^2}E_{n}\right] {\cal Z}(z) = 0,
\end{equation}
recall that the radius of the ring was chosen to be $R=10$ nm and $m$ is the mass of the electron interacting on the axis of symmetry of the charged ring $+100~e$ so that $V_{0} = 14,4$ eV.
For the Numerov numerical method we make the following definitions:
\begin{eqnarray}
 \alpha &=& \frac{2m}{\hbar^2}{V_0} = 377,955306845329/(\mbox{eV nm}^2), \cr\cr
 \beta &=& \frac{2m}{\hbar^2}E = 26,2468963087034 ~ E_{n}/(\mbox{eV nm})^2,
\end{eqnarray}
where in the parameter $\beta$ we put the eigenvalue $E_{n}$ variable so that the computational method solves the Schrödinger equation (\ref{Schrodinger_8}) and calculate the eigenvalue $E_{n}$.

\begin{figure}[!ht]
%\begin{minipage}[!h]{0.4\linewidth}
\includegraphics[width=1.0\linewidth]{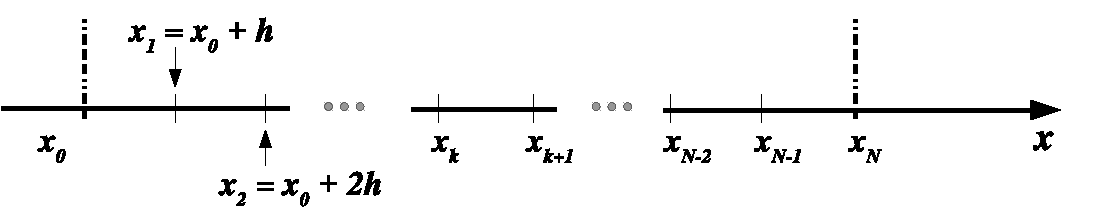}%\vspace{1.0cm}
\caption{We assume that the range is discretized into $N$ equal parts: $h = x_{k+1} - x_{k}$. }
\label{eixo_x}
%\end{minipage}
\end{figure}

The numerical method of Numerov solves differential equations in the format below:
\begin{equation}
 \label{Numerov_1}
 \frac{d^2 Y(x)}{dx^2} = - g(x)Y(x) + s(x).
\end{equation}
Applying the Schrödinger equation to the above equation format (\ref{Numerov_1}), making the respective changes of variables: ${\cal Z}(z) \rightarrow Y(x)$, it results that $s(x) = 0$ and
\begin{equation}
\label{Numerov_2}
 g(x) = \frac{\alpha}{ \sqrt{1+\frac{x^2}{R^2}}} + \beta E_{n}.
\end{equation}
We have segmented the range of the variable $x$ into $N$ values as highlighted in the diagram in Figure  \ref{eixo_x}.
\\
The Numerov method algorithm states that,
\begin{equation}
 \label{Numerov_3}
 Y_{k+1} = \frac{2Y_{k} \cdot \left(1-\frac{5h^2}{12} ~ g_{k} \right) - Y_{k-1}\cdot \left(1+\frac{h^2}{12} ~ g_{k} \right)}{1 + \frac{h^2}{12} ~ g_{k+1}},
\end{equation}
where the function $g_{k}$ is given by the equation (\ref{Numerov_2}) and that in the algorithm generates values given by
\begin{equation}
 g_{k} =  \frac{\alpha}{ \sqrt{1+\frac{x_{k}^2}{R^2}}} + \beta E_{n}
\end{equation}
where $x_{k} = x_{0} +kh$.

\begin{figure}[h]
%\begin{minipage}[!h]{0.4\linewidth}
\includegraphics[width=1.0\linewidth]{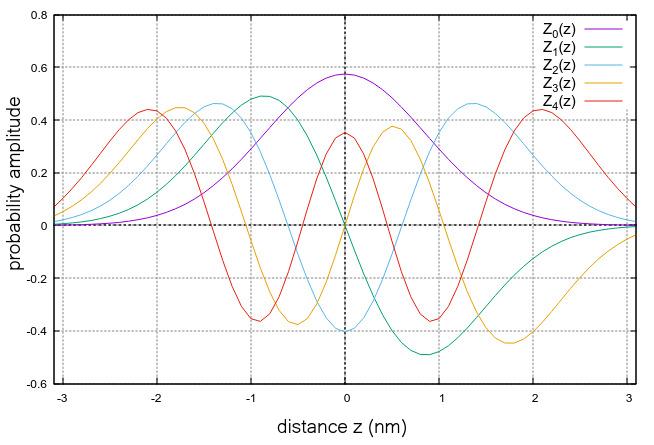}%\vspace{1.0cm}
\caption{Normalized wave functions of the first five eigenstates. Obtained from ordered pairs calculated by Numerov's numerical method.}
\label{autoestados_Numerov}
%\end{minipage}
\end{figure}

To start the iterations we need two starting points $Y_{k-1} = Y(x_{k-1})$ and $Y_{k} = Y(x_{k})$. For $k=1$ in the equation (\ref{Numerov_3}) we get the value of $Y_{2}$ by calculating the values of $Y_{0}$ e $Y_{1}$, and from these points jointed the computer program calculates all possible values of the equation (\ref{Numerov_3})until it ends in $x_{N}$.
It is here in this application of the Numerov method that we highlight the great utility of the Theory of Perturbation method to obtain the necessary initial conditions to numerically solve the differential equation (\ref{Schrodinger_8}).
For example, to obtain the numerical solutions of the first eigenstate, we do $h=0,1$ and we use equation (\ref{autoestado_Z_0}) to find values close to zero of the wave function that are relatively close to the origin of the Cartesian axis.
We calculated and found that for $x_{0} = -3,1$ we have $Y_{0} = 0,001259$ and for $x_{1} = -3,00$ we have $Y_{1} = 0,001888$. Due to the symmetry of the problem (look at the graphs of the functions of the first eigenstates in Figure \ref{autoestados_Z0_Z1_T_Perturbacao}), for $x_{N-1}=3,00$ we have $Y_{N-1} = 0,001888$ and for $x_{N} = 3,1$ we have $Y_{N} = 0,001259$, then, in this routine, the computer program finishes the calculations when these last two Cartesian pairs are reached. The eigenenergy value should vary around the value calculated approximately by the perturbation theory method.
In Table \ref{tabela2}, the value of energy for $n=0$ is  $E_{0} = -14,3478480690$ eV, then we establish an interval that contains this value. The computer program executing Numerov's numerical method adjusts the eigenfunction that converges to the above initial conditions and at the same time calculates the value of the eigenenergy. This same iterative process is performed for the next eigenstates $n=1,2,3 \cdots$. In this way we present the numerical results for the energies in the Table \ref{tabela3}, which can be compared with the numerical results obtained by the Theory of Perturbation in Table \ref{tabela2}, and the graphs of the respective first five eigenstates in Figure \ref{autoestados_Numerov}.
%
%\newpage

\begin{table}[h]
%\begin{ruledtabular}
	\centering
	\caption{The first five energy levels of the eigenstates of the Schrödinger equation (\ref{Schrodinger_6}), by Numerov's numerical method.}
	\begin{tabular}{|c|c|} \hline \hline
	
	 n  & Energy  (eV) \\
	 \hline
	0 & -14,3478377998
 \\
	\hline
	1 & -14,2439376895
 \\
	\hline
	2 & -14,1408538542
 \\
	\hline
	3 & -14,0383114290
 \\
	\hline
	 4 & -13,9350535238
\\ \hline \hline
	\end{tabular}
		\label{tabela3}
	%\end{ruledtabular}	
	\end{table}

The program to implement Numerov's numerical method was developed in gcc - GNU project C and C++ compiler Versão 12.1.0 Pattern C++11.
Code available at https://github.com/brunocmo/NumerovMethod

We can recall  equation (\ref{energia_anel_1}) where the total energy of the quantum system is indexed with three quantum numbers, $n = 0,1,2, \cdots$, due to the electrostatic potential of the ring, $p=1,2,\cdots$, referring to the zeros of the Bessel harmonic functions due to the circular insurmountable potential barrier of the ring and $\nu=0,1,2\cdots $, with respect to angular momentum. The total energy is then given by,
\begin{equation}
 E_{p\nu n} = \frac{\hbar^2 k_{p\nu}^2}{2m} + E_{n} = E_{p\nu} + E_{n}.
\end{equation}
Using the first value $k_{p\nu}$ highlighted in the equation (\ref{valores_de_k}), we calculate that,
$E_{10} = 0,0022033843$ eV and the ground state energy of the quantum system is given by,
\begin{equation}
 E_{100} = -14,3456344155~\mbox{eV}.
\end{equation}
The energy value of any excited state can be calculated from the numerical results obtained in this section V and in section II.
%

%\newpage

\section{Conclusion}
\indent

We consider a system of an electrostatically charged ring-shaped quantum nanostructure in which a charged quantum particle interacts on the ring symmetry axis. In principle, this structure has physical similarities with the already known nanostructure of the quantum corral whose radial solutions are similar.
Just as the nanostructure of the quantum corral is possible to be experimentally realized, it may be possible to realize the electrostatically charged ring at quantum levels.
The quantum solution on the axis of symmetry $z$ results in the Schrödinger equation (\ref{Schrodinger_5}), and from this problem we show a numerical solution with the method of the Perturbation Theory. We also checked the values of the first five eigenstates with the Numerov method. We emphasize that the Perturbation Theory method helps us a lot as a guide to find the initial conditions that are very useful for the Numerov method to be applied. We conclude that Numerov method is relatively easy to apply in computer programs and can be well used to refine numerical results of a given quantum problem. Finally, we trust that the results presented help scientists and engineers to construct experiments and technological devices.

%\newpage


\begin{thebibliography}{99}


\bibitem{Kittel} C. Kittel, {\it Introduction to Solid State Physics},  John Wiley \& Sons, 8th ed. (2004).

\bibitem{Griffths} D. Griffths, {\it Introduction to Quantum Mechanics}, Addison-Wesley Professional, 2nd Revised ed. (2004).

\bibitem{barticevic} Z. Barticevic, M. Pacheco, A. Latg\'e, Phys. Rev. B 62 (2000) 6963.

\bibitem{chakraborty} T. Chakraborty, P. Pietilainen, Phys. Rev. B 52(3) (1995) 1932.

\bibitem{Danilevich} A. Danilevich, {\it 1D Quantum Rings}, available in https://www.studocu.com/en-us/document/boston-college/quantum-physics/anatoly-danilevich-text/11567312 (2007).

\bibitem{Fomin} V. M. Fomin, {\it Physics of Quantum Rings}, Springer-Verlag Berlin Heidelberg (2014).


\bibitem{browmick} S. Browmick, et al, Appl. Phys. Lett. 96  (2010) 231103.

\bibitem{huang} G. A. Huang, Appl. Phys. Lett. 94 (2009) 101115.

\bibitem{wu} J. Wu, et al, Appl. Phys. Lett. 101 (2012) 043904.

\bibitem{suarez} F. Su\'arez, et al, Nanotecnology 15 (2004) 5126.

\bibitem{neill} C. Neill, et al, Nature 594 (2021) 508.

\bibitem{jellal} A. Belouad, A. Jellal, H. Bahlouli, Eur. Phys. J. B 94 (2021) 75.

\bibitem{Eigler1} M. F. Crommie, C. P. Lutz \& D. M. Eigler, Science, 262 (1993) 218.

\bibitem{Eigler2}  M. F. Crommie, C. P. Lutz \& D. M. Eigler, Nature, 363 (1993) 524.

\bibitem{Eigler3} M. F. Crommie, C. P. Lutz, D. M. Eigler \& E. J. Heller,  Physica D: Nonlinear Phenomena 83 (1995) 98.

\bibitem{nisenet} https://www.nisenet.org/catalog/scientific-image-quantum-corral-top-view

\bibitem{Goswami} A. Goswami, {\it Quantum Mechanics}, Waveland Press Inc., second edition, (1997).

\bibitem{Arfken} G. B. Arfken, J. Webber, {\it Mathematical Methods for Physicists}, fourth edition, Academic Press, (1995).

\bibitem{Wolfram} https://mathworld.wolfram.com/BesselFunctionZeros.html

\bibitem{Jackson} J. D.Jackson, {\it Classical Electrodynamics}, 3rd Edition, published by Wiley (1998).

\bibitem{Sears} H. D. Young, R. A. Freedman, {\it Sears and Zemansky's University Physics with Modern Physics}, 14th edition, Pearson Education, Inc. (2016).

\bibitem{Cohen} C. Cohen-Tannoudji, B. Diu, F. Laloë, {\it Quantum Mechanics, Volume II}, John Wiley \& Sons, (1977).

\bibitem{Wolfram2} Wolfram Alpha computational intelligence, wolframalpha.com

\bibitem {Hairer} E. Hairer, S. P. Nørsett, G. Wanner, {\it Solving Ordinary Differential Equations I - Nonstiff Problems}, Second Revised Edition, Springer-Verlag Berlin Heidelberg (1993).

\bibitem{Caruso} F. Caruso , V. Oguri, Rev. Bras. Ens. Fis. 36 (2)(2014) 2310.

\

\end{thebibliography}
\end{document}